\documentclass[natbib]{svcyclop}
\bibpunct{[}{]}{;}{n}{}{,}
\usepackage{graphicx}    

\begin{document}

\title{Exponential Random Graph Models}

\author{Agata Fronczak}
\institute{Faculty of Physics, Warsaw University of Technology\\
Koszykowa 75, PL-00-662 Warsaw, Poland\\
E-mail: agatka@if.pw.edu.pl}


\maketitle

\section{Synonyms}
p* models, p-star models, p1 models, exponential family of random graphs, maximum entropy random networks, logit models, Markov graphs

\section{Glossary}
\begin{itemize}
\item\textbf{Graph and network}: the terms are used interchangeably in this essay.
\item\textbf{Real-world network}: (real network, observed network) means network data the researcher has collected and is interested in modelling.
\item \textbf{Ensemble of graphs}: means the set of all possible graphs (network realizations) that the (real-world) network may reasonably be expected to become, with the assigned probability distribution, which specifies how likely it is that the network will be found in a particular realization. In other words, ensemble of graphs is defined by ascribing a statistical weight to every graph in the given set.
\item\textbf{Graph observable}: measurable property of a graph.
\item \textbf{Network Hamiltonian}: is a particular type of objective (fitness) function, $H(G)$. The exponential random graph model defines a probability distribution over a specified set of possible graphs, $\mathcal{G}=\{G\}$, such that the probability $P(G)$ of a particular graph $G$ is proportional to $e^{H(G)}$, where $H(G)=\sum_i\theta_ix_i(G)$. In the Hamiltonian, $\{x_i\}$ is the set of graph observables upon which the relevant constraints act, and $\{\theta_i\}$ is a set of ensemble parameters which we can vary so as to match the properties of the model network to the real-world network under investigation.
\item\textbf{Adjacency matrix}: is a matrix with rows and columns labelled by graph vertices $i$ and $j$, with elements $A_{ij}=1$ or $0$ according to whether the vertices, $i$ and $j$, are connected/adjacent or not. In the case of an undirected graph with no self-loops or multiple edges (the so-called \textbf{simple graph}), the adjacency matrix is symmetric (i.e. $A_{ij}=A_{ji}$) and has $0$s on the diagonal (i.e. $A_{ii}=0$). Accordingly, for a \textbf{simple directed graph} the symmetry condition may not be fulfilled, i.e. it can be that $A_{ij}\neq A_{ji}$.
\item \textbf{Reciprocity}: describes tendency of vertex pairs to form mutual directed connections between each other.
\item\textbf{Clustering}: describes tendency of nodes to cluster together. Clustering is measured by the \textbf{clustering coefficient} which calculates the average probability that two neighbors of a vertex are themselves nearest neighbors.

\end{itemize}

\section{Definition}

A graph consists of a set of objects or individuals, called nodes (points, vertices), connected by links (edges). The idea of a graph is a powerful simplification for different phenomena - a way of specifying pairwise relations among a collection of items or agents. Graph models are introduced in order to mimic the patterns of connections in real networks, in an effort to understand the implications of those patterns, or just to describe, how network structures originate, and how they evolve over time.

For example, it has been noted that many networks, including social networks, have degree distributions that roughly follow a power-law: the so-called scale-free networks. A reasonable question would be to ask how the structure of such networks arises, and how the scale-freeness affects their behavior in comparison with the non-scale-free counterparts. To address the first question different models of networks' growth and evolution have been introduced, such as the famous Barab\'{a}si-Albert model of preferential attachment. The second question was/is often addressed with the help of exponential random graph (ERG) models, which are discussed in this essay. The ERG model is particularly useful when one wants to create model networks that match the properties of observed networks as closely as possible, but without going into details of the specific process underlying network formation. Such model graphs are not only interesting in their own right for the light they shed on the structural properties of networks. They can be used to study models of processes taking place on networks, such as epidemics spreading, diffusion of information, or opinion formation in social networks.

Nowadays, exponential random graphs (ERGs) are among the most widely-studied network models.
Different analytical and numerical techniques for ERG have been developed that resulted in the well-established theory with true predictive power. Unfortunately, these advantages come at a price: ERG model is mathematically and conceptually rather sophisticated, and its understanding demands some effort of the reader.

\section{Introduction}

Many of the networks we observe in the real world exist in only one realization (instantiation) that we can study. There is only one Internet, only one World Trade Web, and only one network of social ties formed, for example, between filmmakers through their past collaboration in film projects. Only one realization of a given network does not mean, however, that this concrete realization is the only possible that the network may have. Common sense suggests that in other circumstances different link configurations in the considered networks could arise. In particular, the Internet evolves, therefore one can see its different structural realizations if one looks at different times. By definition, all the snapshots of the Internet are its plausible realizations. It is also reasonable to assume that growing in slightly different conditions (we might say, in a parallel world) the structure of such an alternative Internet would probably have been similar to the real Internet. That is, all plausible realizations of the Internet should have some basic features in common, even if they differ in smaller details. Of course, similar considerations also apply to other types of networks, including economic and financial networks, biological networks, and, obviously, social networks.

Considerations of this kind lead us naturally to the concept of \emph{statistical ensemble of networks}, which is the collection of all possible realizations that the considered network may reasonably be expected to attain, $\mathcal{G}=\{G\}$, plus probability distribution, $P(G)$, over $\mathcal{G}$. In the exponential random graph model each graph, $G$, appears with the probability, $P(G)\propto e^{H(G)}$, that is exponential in the so-called graph \emph{Hamiltonian}, $H(G)$, which determines various networks' properties within the ensemble.

In the following, we show what does it mean "to create" an ensemble of ERGs with a given set of properties, such as a given number of edges, or a given value of the \emph{clustering coefficient} \footnote{The clustering coefficient is a measure of the extent to which nodes in a graph tend to cluster together~\cite{lecturesDorogovtsev}.}. We explain the concept of the network Hamiltonian, which is the key point of ERG theory. Our theoretical derivations are accompanied by example calculations. In particular, the classical random graph model popularized by Erd\"os and R\`enyi is reformulated in the language of ERGs. Other important examples are also discussed, such as the generalized random graphs, the reciprocity model, the so-called two-star model, and Strauss's model of transitive networks. A short training in Monte Carlo simulations, to which the ERG model lends itself admirably, is given. Finally, we also place an emphasis on deep connections of the model with basic principles of equilibrium statistical physics and information theory. In doing so, we argue that these models are not merely an \emph{ad hoc} formulation studied mainly for their mathematical convenience, but a true and correct extension of statistical mechanics to the world of networks.

An excellent basic discussion of exponential random graphs addressed to social science students and researchers is given in \cite{1999Anderson, 2007Robins}. This essay is intentionally designed to be more theoretical in comparison with the well-known primers just mentioned. Given the interdisciplinary character of the new emerging science of complex networks \cite{bookBarabasi}, the essay aims to give a contribution upon which network scientists and practitioners, who represent different research areas, could build a common area of understanding.

\section{Historical Background}

The first truly general ensemble model of networks was introduced by Solomonoff and Rapoport in 1951 \cite{1951Solomonoff}, who considered the collection of all undirected simple graphs with a fixed number of vertices, $N$, in which every pair of nodes was connected with an edge with probability $p$. In the late 1950s and early 1960s, the model was fairly extensively studied by Erd\"{o}s and R\`{e}nyi, see e.g.~\cite{1959ER,1960ER}. Ever since it is known as Bernoulli model or Erd\"{o}s-R\`{e}nyi model. We mention this because this particular ensemble of graphs was indeed the first example of the ERG model, and we will meet it again when we will be discussing concrete examples of exponential random graphs.

The exponential random graph model, as we discuss it in this essay, was first proposed in the early 1980s by Holland and Leinhardt \cite{1981Holland}, who built on statistical foundations laid by Bessag \cite{1974Besag}. Substantial further developments were made by Frank and Strauss \cite{1986FrankStrauss,1986Strauss}, and continued to be made by other authors throughout 1990s \cite{1999Anderson}. In recent years, a number of physicists have also made theoretical studies in the field, see e.g.~\cite{2001Burda, 2002Berg, 2003Burda, 2004aPark, 2006Fronczak, bookNewman}.

Nowadays, exponential random graph models are in common use within the social network analysis (SNA) community \cite{1999Anderson, 2007Robins}. Furthermore, the tool box of standard network models/methods is more frequently equipped with the ERG model. Most likely, it happens because ERGs are perceived as being a practical tool for modelling any complex networks, especially that several standard computer tools are available for simulating and manipulating them, such as the ERGM package \cite{ERGMtool}.

\section{Exponential Random Graphs: Elements of the Theory}

\subsection{Definition of the Model: Graph Hamiltonian and Ensemble Parameters}

Suppose, we have a real-world network and we want to create its ensemble model. For a start, we have to define \emph{graph observables} to be measurable properties of the network, that we want to be reflected in the model. Examples of such observables are: the number of edges, $E$, the degree sequence, $\{k_i\}=k_1,k_2,\dots k_N$, the average shortest path length, $l$, and the clustering coefficient, $C$. In this section, for the purpose of convenience and in order to perform general calculations, we assume that the observables, on which we focuss, are:  $\{x_i\}=x_1,x_2,\dots x_r$, and their values measured in the considered real network are respectively equal to: $\{x_i^*\}=x_1^*,x_2^*,\dots x_r^*$.

\begin{figure}
\centering
\includegraphics[height=5.5cm]{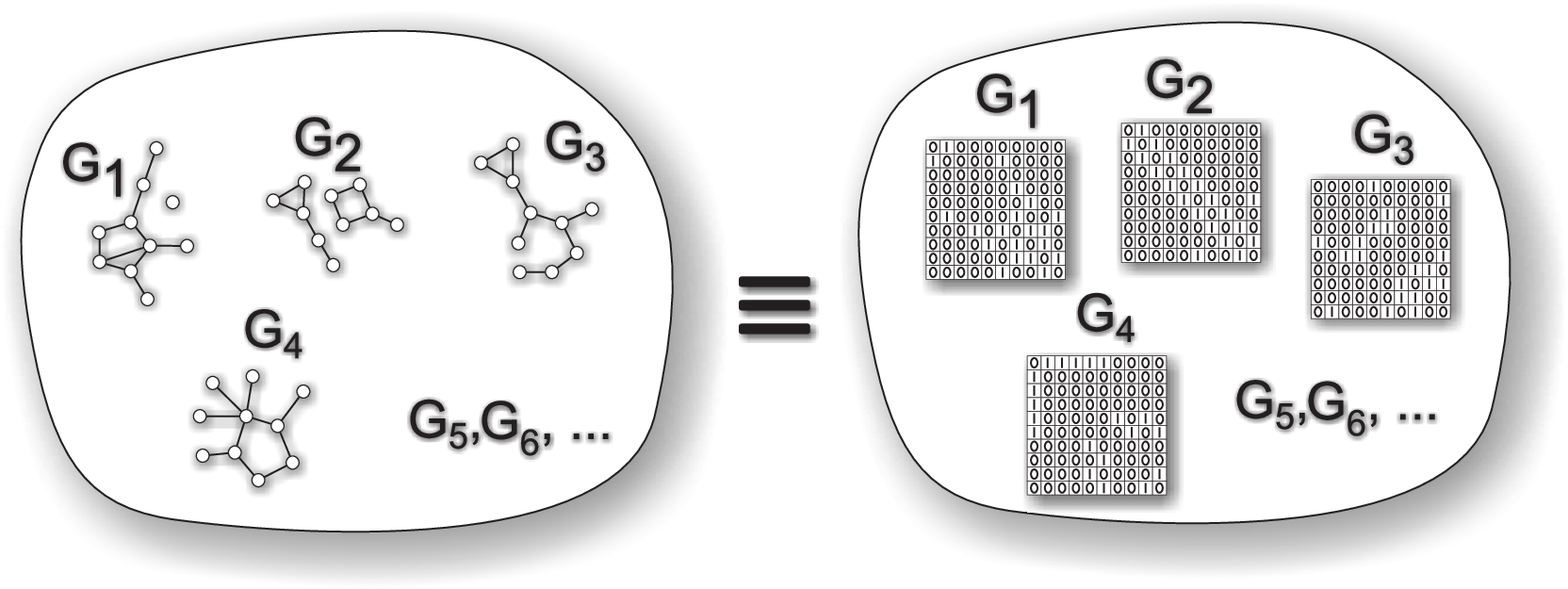}
\caption{Statistical ensemble of simple graphs with a fixed number od nodes.}
\label{fig1}
\end{figure}

Now, we have to specify the collection of all possible realizations that our real-world network may reasonably be expected to attain. It means, we have to define the set of graphs, $\mathcal{G}=\{G\}$, that we want to study. In the following, if not stated otherwise, we restrict ourselves to \emph{simple graphs} with a fixed number of nodes $N$. A simple graph has, at most, one link between any pair of vertices and it does not contain self-loops connecting vertices to themselves. There exists a one-to-one correspondence between simple graphs and symmetric matrices of size $N$ with elements $A_{ij}$ equal to either $0$ or $1$ (see Fig.~\ref{fig1}). If we know that the network has $N$ nodes and there is no reason to think that $N$ will change (or change significantly), and if we know that the direction of the edges does not matter, then the set of simple graphs is a sensible choice for $\mathcal{G}$.

Thus, we have measurements of one or more graph observables and we have also specified plausible realizations of the network. The aim is to choose probability distribution $P(G)$ over the set of all possible network realizations, $\mathcal{G}=\{G\}$. In the exponential random graph model one assumes that the probability distribution, $P(G)$, has the form \footnote{The question, of making the best choice of a probability distribution, $P(G)$, given only a relatively small number of constraints on that distribution, is addressed in the section \emph{Connections with Information Theory and Statistical Physics}.}
\begin{equation}\label{PG0}
P(G)=\frac{e^{H(G)}}{Z},
\end{equation}
where $Z$ is called the partition function and can be calculated form the normalization condition
\begin{equation}\label{Z0a}
\sum_{G\in\mathcal{G}}P(G)=\frac{1}{Z}\sum_{G\in\mathcal{G}}e^{H(G)}=1,
\end{equation}
which implies
\begin{equation}\label{Z0b}
Z=\sum_{G\in\mathcal{G}}e^{H(G)}.
\end{equation}
The network Hamiltonian in Eq.~(\ref{PG0}) is given by
\begin{equation}\label{H0}
H(G)=\sum_{i=1}^r\theta_ix_i(G),
\end{equation}
where $\{x_i(G)\}$ are the values of the observables $\{x_i\}$ for a graph $G$, while $\{\theta_i\}=\theta_1,\theta_2,\dots\theta_r$ are ensemble parameters, which are said to be conjugated to observables in such a way that the value, $x_i^*$, as measured in the real network is equal to the corresponding average value within the ensemble, i.e.
\begin{equation}\label{meanxi}
\langle x_i\rangle=\sum_{G\in\mathcal{G}}x_i(G)P(G)=x_i^*,
\end{equation}
for $i=1,2,\dots r$.

Now, we are in a position to answer the question: What does it mean to define an ensemble of exponential random graphs? In short, the ensemble is defined if one specifies both: plausible network realizations, $\mathcal{G}=\{G\}$, and the observables, $\{x_i\}$, which appear in the network Hamiltonian, see~Eq.~(\ref{H0}). The ensemble parameters, $\{\theta_i\}$, used in the Hamiltonian are calculated (analytically or numerically) from the ensemble constraints \footnote{The constraints, Eq.~(\ref{meanxi}), consist of $r$ equations in $r$ unknowns.} given by Eq.~(\ref{meanxi}).

\subsection{Properties of the Model: Expectation Values and Their Fluctuations}

Once the probability distribution $P(G)$ over $\mathcal{G}$ is given, see~Eq.~(\ref{PG0}), we can use it to calculate estimates of other quantities of interest within the ensemble. For example, the expectation value of a quantity $y$ is given by
\begin{equation}\label{meany}
\langle y\rangle=\sum_{G\in\mathcal{G}}y(G)P(G) =\frac{1}{Z}\sum_{G\in\mathcal{G}}y(G)e^{H(G)}.
\end{equation}
This expression provides the best-guess at the value of the quantity $y$ given the only general constraints, Eq.~(\ref{meanxi}). In other words, the ERG model enables us to answer questions of the form: If we know certain things about a real network, e.g. its measured estimates $\{x_i^*\}$, what is the best estimate for $y$? For example, if we know the average node degree in a network, what is the best estimate for the clustering coefficient? The exponential random graph model gives a rigorous answer to questions of this kind.

An interesting special case arises when the quantity $y$ is itself one of the primary network observables, e.g. $y=x_j$. One may ask: Why we would want to do this, given that, since $x_j^*$  is used as an input to our model, cf.~Eq.~(\ref{meanxi}), we already know its expectation value. The answer is that, to define the ensemble we need to fix the parameters $\{\theta_i\}$. We can do this by calculating expectation values $\{\langle x_i\rangle\}$ for given ensemble parameters, $\{\theta_i\}$, and then varying the parameters until $\{\langle x_i\rangle\}$ take the desired values.

The value of $\langle x_j\rangle$ within the considered ensemble is given by
\begin{eqnarray}\label{meanxj}
\langle x_j\rangle&=&\frac{1}{Z}\sum_{G\in\mathcal{G}}x_j(G)e^{\sum_{i=1}^r\theta_ix_i(G)}= \frac{1}{Z}\frac{\partial}{\partial\theta_j}\sum_{G\in\mathcal{G}}e^{\sum_{i=1}^r\theta_ix_i(G)}=\frac{1}{Z}\frac{\partial Z}{\partial\theta_j}=\frac{\partial F}{\partial\theta_j},
\end{eqnarray}
where
\begin{equation}\label{F0}
F=\ln Z
\end{equation}
is called the \emph{free energy}, which is a function of $r$ variables $\{\theta_i\}$ (alike the partition function, $Z$).

Similarly to Eq.~(\ref{meanxj}), one can show that the second derivative of the free energy, $F$, with respect to $\theta_j$ gives the mean square fluctuations of $x_j$, i.e.
\begin{equation}\label{fluctxj}
\langle x_j^2\rangle-\langle x_j\rangle^2=\frac{1}{Z}\frac{\partial^2 Z}{\partial\theta_j^2}-\frac{1}{Z^2}\left(\frac{\partial
Z}{\partial\theta_j}\right)^2=\frac{\partial}{\partial\theta_j}\left(\frac{1}{Z}\frac{\partial
Z}{\partial\theta_j}\right)=\frac{\partial^2 F}{\partial \theta_{j}^{2}}=\frac{\partial \langle x_j\rangle}{\partial\theta_j}.
\end{equation}
In statistical physics, the last expression is known as the \emph{fluctuation-response relation}.  The l.h.s. of Eq.~(\ref{fluctxj}) describes fluctuations in $x_j$, whereas its r.h.s. characterizes susceptibility of the observable to its conjugated ensemble parameter $\theta_j$. The susceptibility is defined as the derivative of $\langle x_j\rangle$ with respect to $\theta_j$, and describes what happens with $\langle x_j\rangle$, when one changes its conjugate parameter $\theta_j$, which determines/represents external conditions related to $x_j$.

\subsection{Connections with Information Theory and Statistical Physics}

Let us define the problem \footnote{The discussion here is not used in the sequel and this section can be omitted at the first reading.}. Thus, let $G$ be a graph in the set of possible network realizations $\mathcal{G}$ and let $P(G)$ be the probability of that graph within the ensemble. We would like to choose $P(G)$ so that the expectation value of each of our graph observables $\{x_i\}=x_1,x_2,\dots x_r$ within the ensemble is equal to its observed value $\{x_i^*\}$. Due to maximum entropy principle of information theory \cite{bookCover}, which amounts to the second law of thermodynamics in statistical physics \cite{bookAttard}, the \emph{best choice} of probability distribution $P(G)$ is the one that maximizes the Shannon/Gibbs entropy,
\begin{equation}\label{S0}
S=-\sum_{G\in\mathcal{G}}P(G)\ln P(G),
\end{equation}
subject to the constraints given by Eqs.~(\ref{Z0a}) and~(\ref{meanxi}).

At this stage, one may ask: What does it mean the \emph{best choice} in this context? In his book on networks (see \cite{bookNewman}, p.~568) Mark Newman explains that the maximum entropy
choice is best in the sense that it makes the minimum assumptions about the distribution
other than those imposed upon us by the constraints. There are choices of distribution we
could make that would satisfy the constraints but would effectively make additional
assumptions. For instance, some choices might make a particular graph or graphs highly
probable while other graphs, only slightly different, are given far lower probabilities. These
would be considered \emph{bad choices} in the sense that they assume things about the ensemble
for which we have no supporting evidence. The entropy as given by Eq.~(\ref{S0}) is precisely a measure of the amount of \emph{assumption} that goes into a particular choice of distribution $P(G)$, or more precisely it is the amount of \emph{antiassumption} or ignorance, and by maximizing it we minimize unjustified assumptions as much as possible.

The maximization of the entropy, Eq.~(\ref{S0}), subject to the constraints of Eqs.~(\ref{Z0a}) and~(\ref{meanxi}), can be done by the method of Lagrange multipliers. Introducing Lagrange multipliers $\alpha$ and $\{\theta_i\}$ one finds that the maximum value of the entropy, $S$, is achieved for the distribution satisfying the following expression
\begin{equation}\label{S1}
\frac{\partial}{\partial P(G)}\left[ S-\alpha\left(1-\sum_{G\in\mathcal{G}}P(G)\right)- \sum_{i=1}^r\theta_i\left(x_i^*-\sum_{G\in\mathcal{G}}x_i(G)P(G)\right)\right]=0,
\end{equation}
for all graphs $G\in\mathcal{G}$. This gives
\begin{equation}\label{S2}
-\ln P(G)-1+\alpha+\sum_{i=1}^r \theta_ix_i(G)=0,
\end{equation}
which implies, cf.~Eq.~(\ref{PG0}),
\begin{equation}\label{S3}
P(G)=\exp\left[\alpha-1+\sum_{i=1}^r\theta_ix_i(G)\right]=\frac{e^{H(G)}}{Z},
\end{equation}
where $H(G)=\sum_{i=1}^r\theta_ix_i(G)$ is the graph structural Hamiltonian, Eg.~(\ref{H0}), and $Z=e^{1-\alpha}$ is the partition function, Eq.~(\ref{Z0b}).

\section{Examples}

In the following we discuss five examples illustrating the ERG model.  The model of \emph{classical random graphs}, which is also known as the Bernoulli model or Erd\"os-R\'enyi model, is the first example. It is the simples ensemble of ERGs, and it can be solved exactly, meaning that one can calculate its partition function and the corresponding derivatives. Next, we discuss \emph{generalized random graphs}, i.e. maximally random networks with a given node degree sequence, which appear to be among the most important network models in the contemporary science of complex networks. Over the past decade, they have been used extensively as a reliable test bed to study dynamical processes on networks. Then the \emph{reciprocity model} is worked out, which is an important example of the historical significance in social network analysis and mining. One says that the idea of exponential random graphs was born along with the model. It is also worth to note that the reciprocity model, like the classical and generalized random graphs, is exactly solvable.

It would be a mistake, however, to think that all or most of exponential random graph models can be exactly solved. Most are not, and to make progress one must turn to approximate analytic solutions (such as the mean field technique) or numerical calculations (such as the Monte Carlo methods). The well-known examples of ERGs, for which only approximate solutions are known, are the \emph{two-star model} and the \emph{Strauss's model of transitive networks}. The two-star model is a toy model. Nevertheless, it is a very important example because it illustrates a crucial property of the exponential family. Namely, the two-star model can produce two radically different classes of networks for the same values of the ensemble parameters \footnote{In the sense of the occurrence probability of a given graph, $P(G)$, it means that the most probable network realizations can be completely different.}. In physics, the feature is known as \emph{spontaneous symmetry breaking}, and it is understood to accompany \emph{phase transitions} which give rise to interesting phenomena, such as ferromagnetism or superconductivity. In the field of social networks, the comprehension of these phenomena is unfortunately rather limited. The symmetry breaking is thought to be, at the least, troubling, although, if presented in the right context, it could be an important/insightful concept used, for example, in the theory of social change.

\subsection{Classical random graphs}\label{ER}

\par Suppose, one knows only the expected number of edges, $\langle E\rangle$, that the undirected network has. Thus, the Hamiltonian of the corresponding ERG model is given by
\begin{equation}\label{Her}
H(G)=\theta E(G),
\end{equation}
where $\theta$ is the ensemble parameter (so-called external field), whose value determines $\langle E\rangle$.

\par In order to proceed, one has to calculate the partition function, $Z$, of the ensemble. When the network Hamiltonian is a simple function of elements $A_{ij}(G)$ of the \emph{adjacency matrix}, as is the case of Eq.~(\ref{Her}) above, where
\begin{equation}\label{Eer}
E(G)=\sum_{i=1}^N\sum_{j=i+1}^NA_{ij}(G)=\frac{1}{2}\sum_{i=1}^N\sum_{j=1(\neq i)}^NA_{ij}(G),
\end{equation}
then the standard way to perform the sum over $\mathcal{G}$ in Eq.~(\ref{Z0b}) is to sum over the elements $A_{ij}$. In the case of simple graphs, the only allowed values for $A_{ij}$ are $0$ and $1$, with $A_{ii}=0$ and $A_{ij}=A_{ji}$. Hence, the resulting partition function underlying the considered ensemble is
\begin{eqnarray}\label{Zer1}
Z&=&\sum_{G\in\mathcal{G}}e^{\theta E(G)}= \sum_{G\in\mathcal{G}}\prod_{i=1}^{N}\prod_{j=i+1}^N e^{\theta A_{ij}}= \prod_{i=1}^{N}\prod_{j=i+1}^N  \sum_{A_{ij}=0}^1e^{\theta A_{ij}}\\\label{Zer2}&=& \prod_{i=1}^{N}\prod_{j=i+1}^N(1+e^\theta)=(1+e^\theta)^{N\choose 2}.
\end{eqnarray}
From this expression one can calculate the free energy, cf. Eq.~(\ref{F0}),
\begin{equation}\label{Fer}
F={N\choose 2}\ln\left[1+e^\theta\right],
\end{equation}
and the average number of edges is the ERG model, cf. Eq.~(\ref{meanxj}),
\begin{equation}\label{Eer1}
\langle E\rangle=\frac{\partial F}{\partial\theta}={N\choose 2}\frac{e^\theta}{1+e^\theta}.
\end{equation}
The last expression can be rearranged to get the value of the only ensemble parameter, $\theta$, behind the given/desired value of $\langle E\rangle$, i.e.
\begin{equation}\label{thetaer}
\theta=\ln\left[\frac{\langle E\rangle}{{N\choose 2}-\langle E\rangle}\right].
\end{equation}

\par At this point, it is insightful to show that the considered ensemble of exponential random graphs is equivalent to the so-called binomial model, which is an alternative definition of the ensemble of classical random graphs, that was famously studied by Erd\"os and R\'enyi. In classical random graphs with $N$ vertices, every pair of nodes is connected with a given probability $p$. Consequently, the total number
of edges is a random variable with the expectation value
\begin{equation}\label{Eer2}
\langle E\rangle={N\choose 2}p,
\end{equation}
Note, that the last expression agrees with Eq.~(\ref{Eer1}) above, when
\begin{equation}\label{per}
p=\frac{e^\theta}{1+e^\theta}.
\end{equation}
Eq.~(\ref{per}) shows that there is a direct relationship between the ensemble parameter, $\theta$, in the ERG model and the connection probability, $p$, in the well-known classical random graphs.

\par The equivalence between the two models becomes even more apparent when looking at the probability of obtaining a graph $G$ (with $N$ nodes and $E$ edges) by the classical random graph construction procedure, which is
\begin{equation}
P(G)=p^E(1-p)^{{N\choose 2}-E},
\end{equation}
and the probability of such a graph within the considered ERG model, cf~Eq.~(\ref{PG0}),
\begin{equation}
P(G)=\frac{e^{H(G)}}{Z}=\frac{e^{\theta E}}{(1+e^{\theta})^{N\choose 2}}= \left(\frac{e^\theta}{1+e^\theta}\right)^E\left(\frac{1}{1+e^\theta}\right)^{{N\choose 2}-E},
\end{equation}
It is easy to see that, given Eq.~(\ref{per}), the two expressions coincide.

\subsection{Generalized random graphs}

\par Despite its general nature and simplicity, the classical random graph model \footnote{which is described in the previous example and shown to be equivalent to the ERG model with an expected number of nodes} turns out to have severe shortcomings as a model of real-world networks. Probably the most important shortcoming is that the expected node degree distribution in classical random graphs, which is the Poisson distribution, significantly differs from the node degree distributions in the majority of real networks. To overcome this problem the so-called generalized random graphs has been proposed, in which degrees of all vertices are drawn from a specified (e.g. power-law/scale-free) degree distribution \cite{2001NewmanPRE}.

Thus, suppose that, rather than measuring the total number of edges in a network, we measure degrees of all the nodes. Let us denote by $k_i$ the degree of a node $i$. The complete set $\{k_i\}=k_1,k_2,\dots k_N$ is called the \emph{node degree sequence} of a network \footnote{Note that, since $\sum_ik_i=2E$, we do not need to specify independently the number of edges}. The ERG model appropriate to such a set of observables has the following Hamiltonian \cite{2004aPark}
\begin{equation}\label{HPk1}
H(G)=\sum_{i=1}^N\theta_ik_i(G ),
\end{equation}
where $\{\theta_i\}$ is the collection of $N$ ensemble parameters (one parameter $\theta_i$ for each node $i$). Noting that $k_i(G)=\sum_{j=1}^NA_{ij}(G)$, Eq.~(\ref{HPk1}) can be rewritten as
\begin{equation}\label{HPk2}
H(G)=\sum_{i=1}^N\sum_{j=1}^N\theta_iA_{ij}(G)=\sum_{i=1}^N\sum_{j=i+1}^N(\theta_i+\theta_j)A_{ij}(G).
\end{equation}
Accordingly, the partition function for this ensemble is given by, cf.~Eqs.~(\ref{Z0b}) and (\ref{Zer1}),
\begin{eqnarray}\label{ZPk1}
Z&=&\sum_{G\in\mathcal{G}}\prod_{i=1}^{N}\prod_{j=i+1}^N e^{(\theta_i+\theta_j) A_{ij}}= \prod_{i=1}^{N}\prod_{j=i+1}^N  \sum_{A_{ij}=0}^1e^{(\theta_i+\theta_j)A_{ij}}\\\label{ZPk2}&=& \prod_{i=1}^{N}\prod_{j=i+1}^N(1+e^{(\theta_i+\theta_j)}),
\end{eqnarray}
and the free energy, Eq.~(\ref{F0}), becomes
\begin{equation}\label{FPk}
F=\sum_{i=1}^N\sum_{j=i+1}^N \ln\left[1+e^{(\theta_i+\theta_j)}\right].
\end{equation}

By differentiating the free energy, Eq.~(\ref{FPk}), with respect to $\theta_j$ one gets the expression for the average degree of the node $j$, see Eq.~(\ref{meanxj}),
\begin{equation}\label{meanki1}
\langle k_j\rangle=\frac{\partial F}{\partial \theta_j}= \sum_{i=1}^N\frac{1}{1+e^{(\theta_i+\theta_j)}}=\sum_{i=1}^Np_{ij},
\end{equation}
where
\begin{equation}\label{pij1}
p_{ij}=\langle A_{ij}\rangle=\frac{1}{1+e^{(\theta_i+\theta_j)}}
\end{equation}
is the connection probability between the nodes $i$ and $j$, i.e. the average value of the element $A_{ij}$ of the adjacency matrix \footnote{ Let us note that Eq.~(\ref{pij1}) can be calculated using Eq.~(\ref{meany}).}.

In sparse networks, the probability of any individual edge is small, i.e. $p_{ij}\ll 1$. In such networks Eq.~(\ref{pij1}) factorizes:
\begin{equation}\label{pij2}
p_{ij}\simeq e^{-(\theta_i+\theta_j)}.
\end{equation}
Inserting the factorized probability into Eq.~(\ref{meanki1}) one gets the following expression for the average connectivity of a node $j$
\begin{equation}\label{meanki2}
\langle k_j\rangle \simeq e^{-\theta_j}\sum_{i=1}^Ne^{-\theta_i}=e^{-\theta_j}\sqrt{\langle k\rangle N},
\end{equation}
where the handshaking lemma \footnote{$\sum_{j=1}^N\langle k_j\rangle=\left(\sum_{j=1}^Ne^{-\theta_j}\right)^2=\langle k\rangle N$} has been used. Then, using Eq.~(\ref{meanki2}), the connection probability can be written as
\begin{equation}\label{pij3}
p_{ij}\simeq \frac{\langle k_i\rangle\langle k_j\rangle}{\langle k\rangle N}.
\end{equation}
The obtained formula for the connection probability between two nodes, $i$ and $j$, is the one which is often encountered in the theoretical description of generalized random graphs.

\subsection{Reciprocity model}

A large number of real-world directed networks, including social networks, display the phenomenon of \emph{reciprocity} \cite{bookWasserman1994,2004GarlaschelliPRL}, i.e. the tendency of vertex pairs to form mutual directed connections between each other ("mutual dyads" in the parlance of social network analysis). In this section we discuss the reciprocity model proposed by Holland and Leinhardt \cite{1981Holland}.

In the model, the set of all possible network realizations is the set of all \emph{simple directed graphs} \footnote{A simple directed graph is a directed graph having no multiple edges or loops.} with a fixed number of nodes, $N$. The network Hamiltonian is given by
\begin{equation}\label{Hr1}
H(G)=\beta E(G)+\gamma R(G),
\end{equation}
where $E$ is the number of all directed edges, $R$ stands for the number of vertex pairs with edges running between them in both directions, while $\beta$ and $\gamma$ are ensemble parameters that can be varied to obtain the desired number of all edges, $\langle E\rangle$, and reciprocated edges, $\langle R\rangle$. Using the elements of the adjacency matrix the Hamiltonian can be written as
\begin{eqnarray}\label{Hr2}
H(G)&=&\beta \sum_{i=1}^N\sum_{j=1}^NA_{ij}(G)+\gamma \sum_{i=1}^N\sum_{j=1}^NA_{ij}(G)A_{ji}(G)\\\label{Hr3}&=& \sum_{i=1}^N\sum_{j=i+1}^N[\beta\left(A_{ij}(G)+A_{ji}(G)\right)+2\gamma A_{ij}(G)A_{ji}(G)],
\end{eqnarray}
where the product $A_{ij}A_{ji}\neq 0$ only when both directed connections exist, i.e. $A_{ij}=1$ and $A_{ji}=1$.

The partition function of the model is given by \footnote{cf. detailed calculations shown in Eqs.~(\ref{Zer1}) $-$ (\ref{Zer2})}
\begin{eqnarray}\label{Zr1}
Z&=&\prod_{i=1}^N\prod_{j=i+1}^N\sum_{A_{ij}=0}^1\sum_{A_{ji}=0}^1e^{\beta(A_{ij}+A_{ji})+2\gamma A_{ij}A_{ji}}=\prod_{i=1}^N\prod_{j=i+1}^N(1+2e^\beta+e^{2(\beta+\gamma)})\\\label{Zr2} &=&(1+2e^\beta+e^{2(\beta+\gamma)})^{N\choose 2},
\end{eqnarray}
and, correspondingly, its free energy is
\begin{equation}\label{Fr}
F={N\choose 2}\ln\left[1+2e^\beta+e^{2(\beta+\gamma)}\right].
\end{equation}

\begin{figure}
\centering
\includegraphics[height=4.5cm]{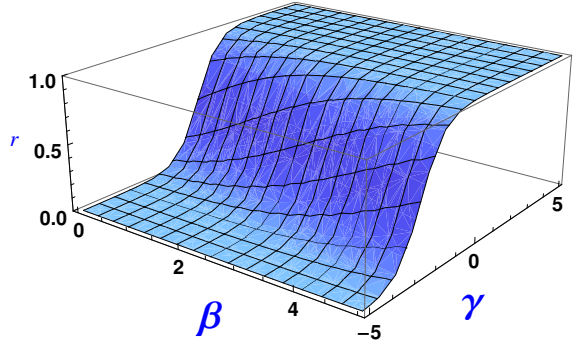}
\caption{The fraction of edges that are reciprocated, Eq.~(\ref{rr}), in the model of Holland and Leinhardt. Note that for varying values of the ensemble parameters, $\beta$ and $\gamma$, the reciprocity parameter, $r$, changes from $0$ (none of edges belongs to mutual dyad) to $1$ (all existing edges are reciprocated).}
\label{fig2}
\end{figure}

Now, using Eq.~(\ref{meanxj}), one finds the expected numbers of all edges and reciprocated edges, i.e.
\begin{equation}\label{meanEr}
\langle E\rangle=\frac{\partial F}{\partial\beta}=2{N\choose 2}\frac{e^\beta+e^{2(\beta+\gamma)}}{1+2e^{\beta}+e^{2(\beta+\gamma)}},
\end{equation}
and
\begin{equation}\label{meanRr}
\langle R\rangle=\frac{\partial F}{\partial\gamma}=2{N\choose 2}\frac{e^{2(\beta+\gamma)}}{1+2e^{\beta}+e^{2(\beta+\gamma)}}.
\end{equation}
One can also calculate the reciprocity parameter, $r$, which is the fraction of edges that are reciprocated:
\begin{equation}\label{rr}
r=\frac{\langle R\rangle}{\langle E\rangle}=\frac{1}{1+e^{-(\beta+2\gamma)}}.
\end{equation}
From Eqs.~(\ref{meanEr}) $-$ (\ref{rr}), it is evident that it is possible to control both the number of edges and the level of reciprocity in the network by suitable choices of the ensemble parameters $\beta$ and $\gamma$ (see Fig.~\ref{fig2}).

\subsection{Two-star model}

\begin{figure}
\centering
\includegraphics[height=4.0cm]{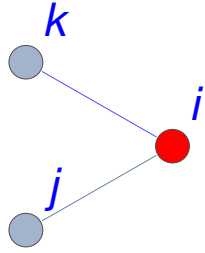}
\caption{A two-star meaning a vertex connected by edges to two other vertices.}
\label{fig3}
\end{figure}

In the two-star model one specifies the expected number of edges, $\langle E\rangle$, and the expected numer of the so-called \emph{two-stars}, $\langle V\rangle$. Possible network realizations correspond to all undirected simple graphs with a given number of nodes, $N$. A two-star is a vertex connected by edges to two other vertices (see Fig.~\ref{fig3}). Varying the number of two-stars allows one to control the extent to which edges in the network stick together, meaning that the edges share common vertices. If the only number of edges in a network is fixed, then sticking together is a random processes, i.e. edges may stick or they may not. On the other hand, if we assume a large number of two-stars, then the edges have to stick together to create the required value of $\langle V\rangle$. In this way, the two-star model allows one to control the extent to which the edges gather together in clumps or are distributed more randomly.

The number of two-stars in a network is
\begin{eqnarray}\label{Vts1}
V(G)&=&\sum_{i=1}^N\sum_{j=1(\neq i)}^N\sum_{k=1(\neq i,j)}^NA_{ij}(G)A_{ik}(G)\\ \label{Vts2}&=& \frac{1}{2}\sum_{i=1}^N\sum_{j=1(\neq i)}^NA_{ij}(G)\sum_{k=1(\neq i,j)}^N \left(A_{ik}(G)+A_{jk}(G)\right),
\end{eqnarray}
and, therefore, the Hamiltonian of the model is given by
\begin{eqnarray}\label{Hts1}
H(G)&=&\beta E(G)+\gamma V(G)\\\label{Hts2}& =& \frac{1}{2}\sum_{i=1}^N\sum_{j=1(\neq i)}^NA_{ij}(G)\left(\beta+\gamma\sum_{k=1(\neq i,j)}^N\left(A_{ik}(G)+A_{jk}(G)\right)\right),
\end{eqnarray}
where $\beta$ and $\gamma$ are two ensemble parameters, and $E(G)$ was already given by Eq.~(\ref{Eer}).

Unfortunately, it is not easy to calculate the partition function, Eq.~(\ref{Z0b}),  for the two-star model. To tell the truth, only approximate solutions of the model are known, such as the mean field solution \cite{2004bPark,bookNewman}. In the following, we present this solution. Those, who are not accustomed to analytical calculations may want to skip them, but we honestly encourage to be persistent because behavior of the two-star model is really crucial. Comprehension of the so-called spontaneous symmetry breaking, that is shown to naturally emerge in this model, can be observed in a number of ERG models, including the famous Strauss's model, which is discussed in a more qualitative (i.e. not a quantitative) way in the next section.

Thus, the two-star model can be solved using the so-called mean-field technique, which is borrowed from statistical physics. First, let us note that the term $\sum_{k=1(\neq i,j)}^NA_{ik}$ in the Hamiltonian, Eq.~(\ref{Hts2}), is simply the number of edges attached to vertex $i$, but excluding the connections to $i$ and $j$. All edges in the model are equivalent \footnote{i.e. they do not have individual properties to distinguish them}, so the average connection probability,
\begin{equation}\label{pts}
p=\langle A_{ik}\rangle,
\end{equation}
is the same for all the pairs of nodes. Therefore, the third sum in Eq.~(\ref{Hts2}) can be written as
\begin{equation}\label{MFts1}
\sum_{k=1(\neq i,j)}^N(A_{ik}(G)+A_{jk}(G))\!\simeq\!\!\sum_{k=1(\neq i,j)}^N(\langle A_{ik}\rangle+\langle A_{jk}\rangle)\!=\!2(N-2)p\!\stackrel{N\gg 1}{\simeq}\!2Np,
\end{equation}
and, correspondingly, the Hamiltonian of the two-star model becomes equivalent to the Hamiltonian underlying classical random graphs, cf.~Eq.~(\ref{Her}),
\begin{eqnarray}\label{Hts3}
H(G)&\simeq &\frac{1}{2}(\beta+2\gamma Np)\sum_{i=1}^N\sum_{j=1(\neq i)}^N A_{ij}(G)\\ \label{Hts4}&=&(\beta+2\gamma Np) E(G)\\\label{Hts5}&=&\theta E(G),
\end{eqnarray}
except that the ensemble parameter in the new Hamiltonian, i.e.
\begin{equation}\label{thetats}
\theta=\beta+2\gamma Np,
\end{equation}
is a rather complicated function of the original parameters, $\beta$ and $\gamma$, and the unknown connection probability, $p$.

Now, exploiting the observed equivalence, one can use Eq.~(\ref{Eer1}) to calculate the average number of edges in the two-star model. Inserting Eq.~(\ref{thetats}) into~(\ref{Eer1}) one gets
\begin{equation}\label{Ets1}
\langle E\rangle={N\choose 2}\frac{1}{1+e^{-(\beta+2\gamma Np)}}.
\end{equation}
Then, equating Eq.~(\ref{Ets1}) to
\begin{equation}\label{Ets2}
\langle E\rangle={N\choose 2}p,
\end{equation}
where $p$, is the connection probability, Eq.~(\ref{pts}), gives a self-consistent equation for the parameter $p$, which characterizes "density" of connections within the network, i.e. \begin{equation}\label{pts1}
p=\frac{1}{1+e^{-(\beta+2\gamma Np)}}=\frac{1}{2}\left[\tanh\left(\frac{1}{2}\beta+\gamma Np\right)+1\right].
\end{equation}
In what follows, for convenience in solving for $p$, we define
\begin{equation}\label{BCts}
B=\frac{1}{2}\beta,\;\;\;\;\;\mbox{and}\;\;\;\;\;C=\frac{1}{2}\gamma N,
\end{equation}
so that Eq.~(\ref{pts1}) becomes
\begin{equation}\label{pts2}
p=\frac{1}{2}\left[\tanh(B+2Cp)+1\right].
\end{equation}

\begin{figure}
\centering
\includegraphics[height=12cm]{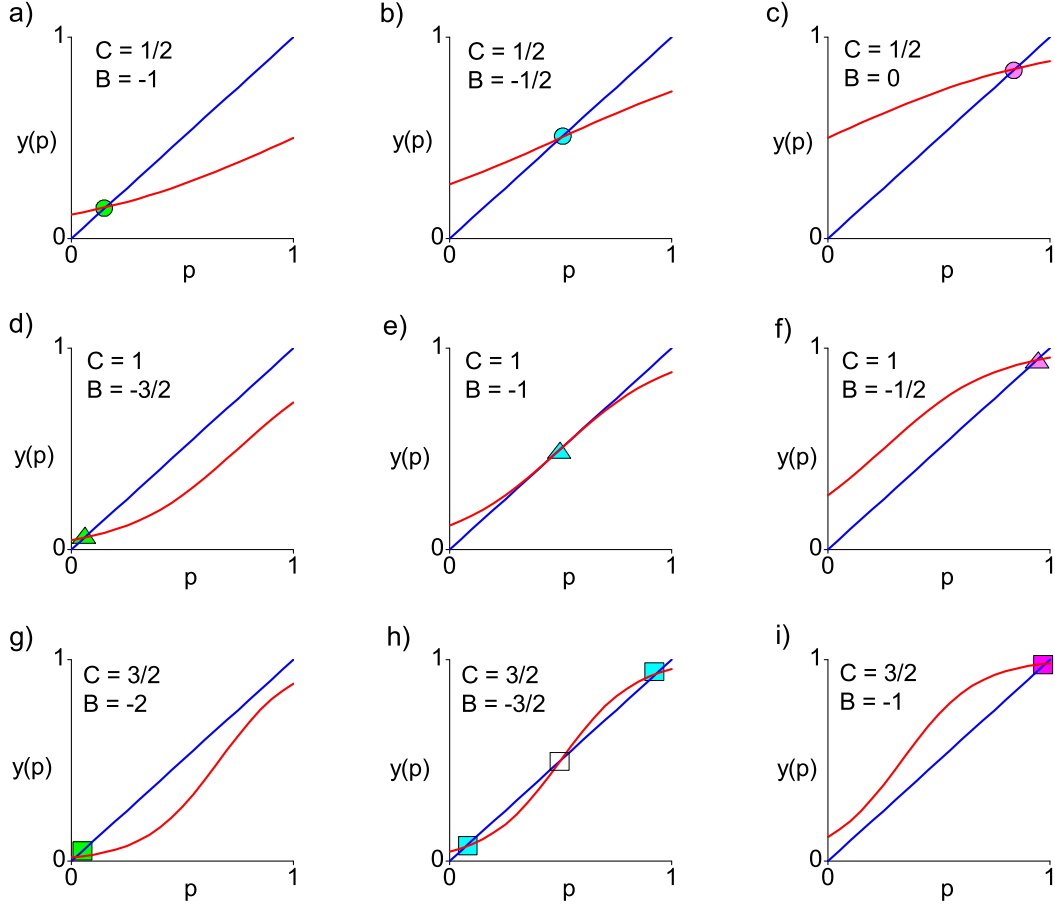}
\caption{Graphical solutions of $p=\frac{1}{2}[\tanh(B+2Cp)+1]$, i.e. Eq.~(\ref{pts2}). Depending on the values of the parameters $B$ and $C$ the line $y=p$ intersects with the curve $y=\frac{1}{2}[\tanh(B+2Cp)+1]$ either three times or only once. Detailed description of the figure is given in the text.}
\label{fig4}
\end{figure}

There is, however, a difficulty with Eq.~(\ref{pts2}): One does not know its closed-form solution. Of course, one can solve this equation numerically, but first it is instructive to see the solutions using a graphical method. If we make plots of the lines $y=p$, l.h.s. of Eq.~(\ref{pts2}), and $y=\frac{1}{2}[\tanh(B+2Cp)+1]$, r.h.s. of Eq.~(\ref{pts2}), as functions of $p$ on the same axis, they will intersect at the solution (or solutions) of the analyzed equation. The method is illustrated in Fig.~\ref{fig4} for different choices of the parameters.

Consider first three plots a),~b), and~c) which show the r.h.s. of Eq.~(\ref{pts2}) (red curves) for $C=\frac{1}{2}$ and three different values of the parameter $B$. Varying $B$ merely shifts the curves horizontally without changing their overall shape. For each curve there is a single point of intersection with the line $y=p$, which is the l.h.s. of Eq.~(\ref{pts2}) (blue lines), indicated by a small circle. As $B$ is varied this intersection point moves smoothly between high and low values of $p$. It means that for $C=\frac{1}{2}$ we can tune the density of the network to any desired value by the parameter $B$.

Now, take a look at three plots g),~h), and~i) in Fig.~\ref{fig4}, which illustrate graphical solutions of Eq.~(\ref{pts2}) for $C=\frac{3}{2}$, and again three different values of $B$. Like in the previous case, varying $B$ shifts the solid curves representing the r.h.s. of Eq.~(\ref{pts2}) horizontally, but now there is an important difference. Due to the higher value of $C$ the shape of the curves has changed. It is steeper in the middle than it was previously, and as a result it is now possible at suitable values of $B$ for the curve to intersect with the line, $y=p$,  not just in one place but in three different places. In this regime, there are three different possible solutions for $p$ for the same values of the model parameters, $B$ and $C$. The solutions are indicated by small squares. In fact, it turns out that the middle solution (open square) is unstable and only the two other are realized in practice. These two solutions, however, correspond to very different network realizations. The first solution (the smaller value of $p$) characterizes sparse networks with a few edges, whereas the second one (the larger value of $p$) describes dense networks with many connections. If one were simulate the two-star model on a computer \footnote{using, for example, Monte Carlo methods}, for the same parameter values one would in this regime sometimes observe a high-density network and sometimes low-density one, depending on initial conditions. In general, one would not be able to predict in advance which of these would occur. Worse yet, for $C=\frac{3}{2}$ there are some values of $p$ that are simply impossible to reach.

This interesting behavior is called spontaneous symmetry breaking. In physics, such a behavior is understood to accompany phase transitions. It is known to give rise to a number of important phenomena which rely on/emerge from collective behavior of the system's constituents. The well-known examples of such phenomena are spontaneous ordering or condensation/nucleation effects in the condensed matter physics. More frequently, however, it happens that researchers from disciplines other than physics strive to use these concepts to understand strange/surprising/unexplained observations. It seems that the borrowed from physics concepts of symmetry breaking and phase transitions may alow understanding properties of systems which are not of interest within the traditional domains of physics, but refer to widely circulated research on complex systems, which becomes more and more influential.
In the SNA community, however, the prevailing view is that such a behavior, that a model can produce two radically different classes of networks for the same values of the model parameters is thought to be confusing/false. On the contrary, from the point of view of the physicists, the behavior is very intriguing/promising. It suggests, that there exist external conditions at which the (social) network becomes very fragile. In the real world, such a fragile/critical network may abruptly change its properties \cite{2007Fronczak}. One says that the network is susceptible to external perturbations. Thus, why not to use the concepts to analyze revolutions, political conflicts, etc.

Finally, for the sake of completeness, one should note that the remaining three plots d),~e) and~f) in Fig.~\ref{fig4} correspond to the critical value of the parameter $C=1$, above which the symmetry breaking appears, and below which it is absent.

\begin{figure}
\centering
\includegraphics[height=8cm]{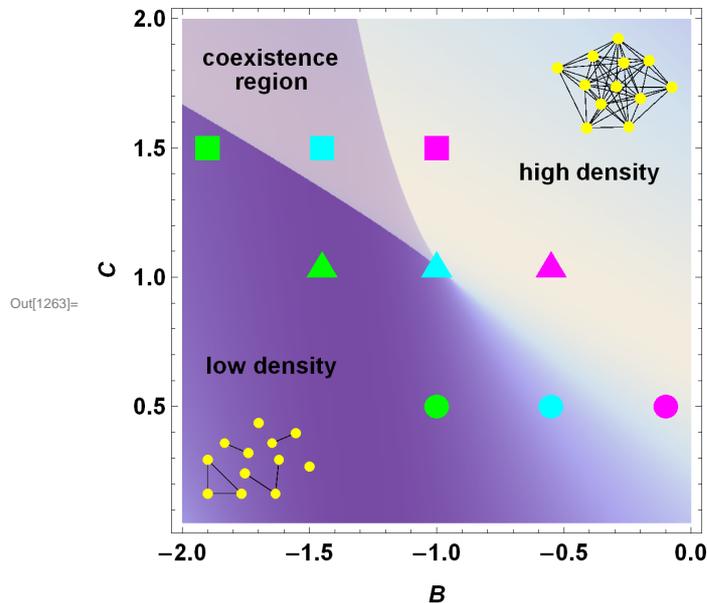}
\caption{Phase diagram of the two-star model in $(B,C)$ space. Detailed description of the figure is given in the text.}
\label{fig5}
\end{figure}

In Fig.~\ref{fig5}, we show the so-called phase-diagram of the two-star model. In the diagram, numerical solutions of Eq.~(\ref{pts2}) are depicted showing different regimes or "phases" of the model as a function of its two parameters, $B$ and $C$. The specific solutions of Eq.~(\ref{pts2}) which are indicated in Fig.~\ref{fig4} by circles, triangles and squares are also adequately plotted in Fig.~\ref{fig5}.

\subsection{Strauss's model of transitive networks}

The famous Strauss model \cite{1986Strauss} was originally proposed as a model of a clustered network. In the model one specifies the expected number of edges, $\langle E\rangle$, and the expected number of triangles \footnote{i.e. cycles of length $3$}, $\langle T\rangle$. Consequently, the network Hamiltonian is given by
\begin{equation}\label{HS}
H(G)=\theta E(G)+\alpha T(G).
\end{equation}

The Strauss model, like the two-star model, can be solved approximately using a mean-field technique \cite{2005Park}. In this case, however, the details of calculation are more complicated then for the two-star model. For this reason, we carry out a short qualitative discussion of the model, instead of quantitative analysis.

Thus, there is symmetry breaking in the Strauss model. Phase diagram of the model resembles the one, that we have already seen for the two-star model. From the diagram, it is evident that there is a structural phase transition in the considered ensemble of networks beyond which the system develops a coexistence region where two distinct classes of networks can be observed. One class corresponds to high density networks, while the other to low density graphs. In this coexistence region, it also happens that no choice of model parameters gives networks of medium density. As a result, there are networks that simply cannot be generated by the model. The observation that in some circumstances the exponential random graph model does not give desired results is a disturbing finding.

\section{Monte Carlo Simulations of ERGs}

The mathematically tractable models are very rare in the exponential family of random graphs \footnote{The solvable examples discussed in the previous section belong to the not very numerous exceptions to the rule}. For this reason, in the absence of analytic progress on various models (Hamiltonians), researchers have turned to Monte Carlo simulations, a numerical method which is ideally suited to exponential random graphs. In what follows we briefly describe this method.

Thus, once the values of ensemble parameters $\{\theta_i\}$ in the Hamiltonian, see Eq.~(\ref{H0}), are specified, the form of the probability distribution $P(G)$, Eq.~(\ref{PG0}), makes generation of graphs correctly sampled from the ensemble straightforward using a Metropolis-Hastings type Markov chain method \cite{2002Snijders, ERGMtool, bookMonteCarlo}. In the method, one defines a move-set in the space of graphs and then repeatedly generates moves from this set, accepting them with probability
\begin{equation}\label{MC1a}
p=1\;\;\;\;\;\mbox{if}\;\;\;\;\;P(G')>P(G),
\end{equation}
where $G'$ is the graph after performance of the move, and
\begin{equation}\label{MC1b}
p=\frac{P(G')}{P(G)}\;\;\;\;\;\mbox{if}\;\;\;\;\;P(G')<P(G),
\end{equation}
while rejecting them with probability $1-p$. Because of the exponential form of $P(G)$, the acceptance probability which is given by Eq.~(\ref{MC1b}) is particularly simple to calculate. It can be written as
\begin{equation}\label{MC2a}
p=e^{H(G')-H(G)}=e^{\Delta H}\;\;\;\;\;\mbox{if}\;\;\;\;\;\Delta H<0,
\end{equation}
where
\begin{equation}\label{MC2b}
\Delta H=\sum_{i=1}^r\theta_i\left(x_i(G')-x_i(G)\right).
\end{equation}
Let us also note, that with the help of $\Delta H$, the condition for certain acceptance of a change, Eq.~(\ref{MC1a}), becomes
\begin{equation}\label{MC2c}
p=1\;\;\;\;\;\mbox{if}\;\;\;\;\;\Delta H>0.
\end{equation}

The choice of the right move-set first and foremost depends on the set of all possible network realizations, $\mathcal{G}=\{G\}$, underlying the studied ensemble. For example, suitable move-sets are: i) addition and removal of edges between randomly chosen vertex pairs for the case of graphs which do not have a fixed number of edges; ii) movement of edges randomly from one place to another for the case of fixed edge numbers but variable degree sequence; iii) edges swaps of the form $\{(v_1,w_1),(v_2,w_2)\}\rightarrow \{(v_1,w_2),(v_2,w_1)\}$ for the case of fixed degree sequence, where $(v_1,w_1)$ denote an  edge from vertex $v_1$ to vertex $w_1$. Monte Carlo numerical simulations of this type are simple to implement and appear to converge quickly allowing one to study quite large graphs.

To be concrete, let us work out the Metropolis algorithm for classical random graphs, i.e. the ERG model with an expected number of edges, cf.~Eq.~(\ref{Her}). The ensemble can be obtained in the following way:
\begin{itemize}
\item[1.] At the beginning one creates any simple graph (i.e. its adjacency matrix) with a given number of nodes, $N$. The starting configuration may be, for instance, the edgeless graph.
\item[2.] Next, in the following time steps, one randomly chooses a matrix element, $A_{ij}(G)$, to be considered for change. For the case, when $A_{ij}(G)=1$ ($0$, respectively) one considers delation (addition) of the edge, i.e. $A_{ij}(G')=0$ ($1$, respectively). This corresponds to the move-set: addition and removal of edges between randomly chosen vertex pairs. Whether the change is accepted depends on $\Delta H$. In classical random graphs, since $H(G)=\theta E(G)$ one has $\Delta H=\pm\theta$ with the upper (lower) sign relating to addition (delation, respectively) of an edge. Therefore, the acceptance criteria, Eqs.~(\ref{MC2a}) and (\ref{MC2c}), depend on the sign of the ensemble parameter $\theta$, which is given by Eq.~(\ref{thetaer}).
\item[3.] The updating of elements $A_{ij}$ should be continued until the network observables stabilise around their mean values \footnote{The mean values result from the ensemble parameters.}. In the case of classical random graphs, the average number of edges, $\langle E\rangle$, should place itself around the value, which is given by Eq.~(\ref{Eer}). Once numerical simulations stabilise, graphs which appear in the course of subsequent updates of the adjacency matrix appear to be correctly sampled network realizations of the studied ensemble.
\end{itemize}

\section{Key Applications and Future Directions}

\subsection{Bimodality and Symmetry Breaking in Social Networks}

In working with exponential random graph models the aim is to create model networks with properties similar to those seen in real-world networks. From the statistical point of view, the ensemble approach seems to be a very logical one. The construction of the model using maximum entropy principle, which is a driving force behind many natural and manmade systems, is natural and ussually gives sensible results. However, the overwhelming opinion within the SNA community is that the two features of ERGs, i.e. spontaneous symmetry breaking, and the ranges of network properties that simply cannot be created using the models, while at the same time real-networks can and do display properties in these ranges, indicate that there is a fundamental flaw or gap in the reasoning behind ERGs.

Strauss himself was already aware of these problems when he proposed his model of transitive networks in the 1980s, and the fact that the problems with Strauss's model are still under active discussion indicates that there is a genuine deadlock here. The issues must be clarified to make progress in the field of ERG models. To shed light on the subject (and, maybe, to start a new debate) let us once again comment on the problems with bimodality and symmetry breaking. The SNA community perceives these features of ERGs as their serious drawbacks. However, symmetry breaking and phase transitions are, in fact, Hamiltonian-dependent inherent  features of maximal entropy random graphs. If one observes these effects in the network model and at the same time one is convinced that the effects are not possible in the real-world network upon which the model is originally based, it means that the considered network Hamiltonian is wrong. Imagine also that one can somehow guess the correct Hamiltonian of the real-world network. Then the ERG model may allow to make reliable predictions about possible/future realizations of this network \cite{2007Fronczak}. It is not possible to cheat maximum entropy principle \footnote{In physics, the principle amounts to the second law of thermodynamics.} and mathematics behind it \footnote{In particular, one should not forget that exponential random graph models can be only used to describe the so-called equilibrium networks. The (real-world) equilibrium network is the one, which is not evolving quickly and, therefore, its properties do not change abruptly.}.

Thus, maybe the problem is not with the ERG model itself, but stems from inadequate construction of networks' Hamiltonians? In particular, let us look at the Hamiltonian of Strauss's model, Eq.~(\ref{HS}). Is it adequate to describe social networks? In view of the findings of the science of complex networks, it is not. A major problem with this model is that its Hamiltonian is designed for homogeneous systems, in which edges and triangles are uniformly distributed over the network. Unfortunately, such a homogeneity is not observed in real-world social networks, which are strongly inhomogeneous in terms of the local connectivity and the local clustering \cite{bookBarabasi, lecturesDorogovtsev}. Therefore, it seems that the simples way to overcome the problematic issues with Strauss's model is just to refine its Hamiltonian, because mathematics behind the model is correct \cite{2005Park}.



\subsection{Predictability and Prediction}

One says that the ultimate proof of our understanding of natural and manmade systems is reflected in our ability to predict/controll them. In statistical physics, the problem of predictability/controllability is tackled with the help of fluctuation/stability theory and response theory - the two pillars on which non-equilibrium statistical mechanics is built.

In physics it is well-understood that interactions between individuals in a system often lead to an emergent global behavior. This behavior is not only imposed by external controllers/conditions, but may result from internal interactions/details. These issues are closely related to the problem of possible microscopic realizations of the system \footnote{This is in fact the so-called ensemble formulation of statistical physics \cite{bookAttard}.}, their macroscopic perception in terms of the system's phases, and the corresponding phase transitions (or symmetry breaking phenomena). Furthermore, given the ensemble approach to random graphs described in this essay, which directly corresponds to ensemble formulation of statistical physics, one can use and modify the well-known methods from physics (such as fluctuation-response relations, and Onsager reciprocal relations) to apply them to inference on, and analysis of exponential random graph models.

We already know that such generalizations can be done. To be concrete, the ideas have been already applied in purely theoretical considerations about ERGs \cite{2006Fronczak}. Recently, one has also used them to discuss properties and predictability of real-world networks. In particular, having the mathematically tractable yet realistic ERG model of the world trade network, one has shown that bilateral trade fulfils a simple fluctuation-response theorem describing the susceptibility of trade volume to changes in gross domestic products of trade partners \cite{2012aFronczak,2012bFronczak}. Therefore, one can also believe that the analogous quantitative response-like relations may also characterize social networks allowing, for example, to predict under which conditions the connected network may disintegrate into disconnected clusters.

\section{Cross-References}
\begin{itemize}
\item[1.] Models of Social Networks
\item[2.] Network Models
\item[3.] Scale-free Nature of Social Networks
\end{itemize}

\section{Acknowledgements}

A.F. acknowledges financial support from the Ministry of Science and Higher Eduction in Poland (national three-year scholarship for outstanding young scientists, 2010-2013) and from the Foundation for Polish Science (grant no. POMOST/2012-5/5).

\bibliographystyle{spbasic}


\end{document}